\documentclass[aps,prl,groupedaddress,twocolumn,showpacs]{revtex4-1}

\usepackage{graphicx}
\usepackage{bm}
\begin{document}


\title{Magnetic phase diagram and Mott transition of the half-filled 1/5-depleted Hubbard model with frustration}


\author{Atsushi Yamada}
\email[]{atsushi@physics.s.chiba-u.ac.jp}
\affiliation{Department of Physics, Chiba University, Chiba 263-8522, Japan}

\date{\today}

\begin{abstract}
The magnetic properties and Mott transition of the half-filled Hubbard model on the 1/5 depleted square lattice with frustration 
is studied at zero temperature by the variational cluster approximation. 
The $(\pi,\pi)$ N\'eel ordering (AF) is stable in a wide region of the phase diagram 
and almost completely veils the non-magnetic Mott transition for the unfrustrated case. 
However, AF is severely suppressed by the frustration and even with moderate frustrations 
the non-magnetic Mott transition takes place in the range where the intra-dimer hoppings are larger than the 
intra-plaquette hoppings.

\end{abstract}




\maketitle

\begin{figure}[t]
\begin{center}
\includegraphics[width=8.6cm,bb= 122 320 529 582]{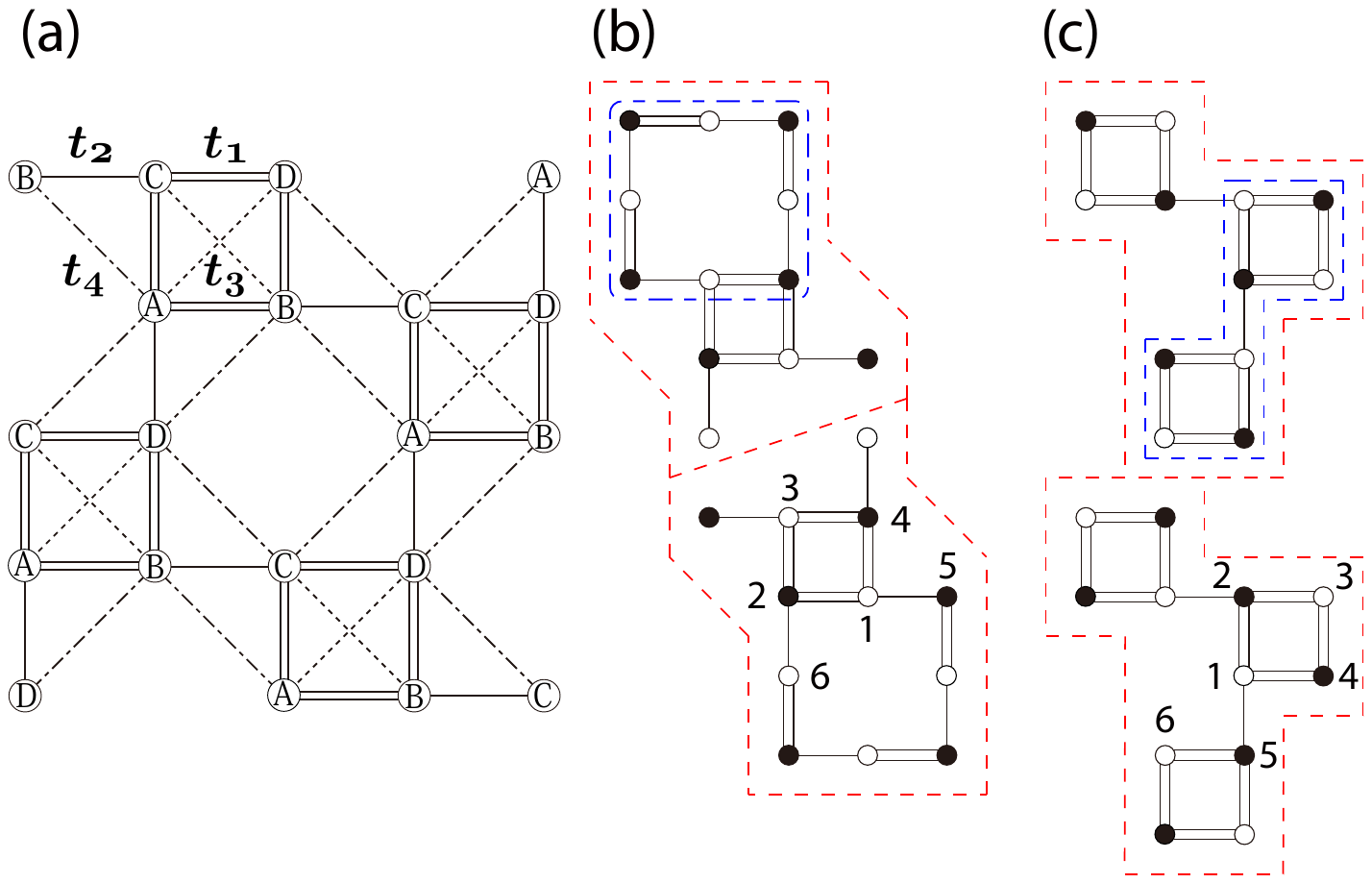}
\caption{
(Color online) (a) Schematic view of the hopping terms $t_1 \sim t_4$ in the 1/5-depleted square lattice.
(b)-(c) The 12-site (the dashed red lines) and 8-site (dash-dotted blue lines) clusters 
used in VCA. The filled (unfilled) circles correspond to up (down) spins in the $(\pi,\pi)$ N\'eel ordering. 
The 12-site clusters are mirrored to recover the lattice geometry. 
\label{fig:model}}
\end{center}
\end{figure}

{\it Introduction.}--- When the kinetic and Coulomb repulsion energies are competing,  
low-dimensional materials exhibit a variety of phenomena like superconductivity, Mott transition, and spin liquid, depending 
on their lattice structures and degrees of the geometric frustrations. 
One of the interesting lattice structures is the 1/5-depleted square lattice (See Fig. \ref{fig:model} (a)) 
realized e.g. by CaV$_4$O$_9$\cite{taniguchi} and iron selenide families\cite{iron-selenid}. 
This is a non-Brave lattice and its multi-band structure will provide greater flexibility in 
controlling the electronic properties of materials of this lattice structure. 
In the limit of the infinite Coulomb repulsion, the 
$(\pi,\pi)$ N\'eel ordering (See Fig. \ref{fig:model} (b) and (c)) separates two spin gapped states, 
one adiabatically connected to decoupled dimers, the other connected to decoupled plaquettes, by the 
second order quantum transition~\cite{katoh,ueda,troyer,troyer2,starykh,gelfand,albrecht}. 
As for the effect of the electron correlations, the Mott transition 
was recently studied using the Hubbard model 
by cluster dynamical mean-field theory(CDMF)~\cite{yanagi}, 
but this analysis assumed the non-magnetic ground state and 
if the Mott transition is veiled or not by the $(\pi,\pi)$ N\'eel ordering is not clear. 
The magnetic properties are investigated by the determinant quantum Monte Carlo(DQCM)\cite{khatami}, 
which reports that the dominant magnetism is the $(\pi,\pi)$ N\'eel ordering, but the Mott transition was not analyzed. 
Moreover, the effect of the next nearest hoppings, which is in general not negligible\cite{starykh,gelfand} 
and expected to largely affect the magnetic properties, was not considered in these analyses\cite{yanagi,khatami}. 
In this paper we investigate the magnetic phase diagram and Mott transition in 
the half-filled Hubbard model on the 1/5-depleted square lattice taking into account the effect of the frustration 
at zero temperature by the variational cluster approximation (VCA). 

We found that the $(\pi,\pi)$ N\'eel ordering (AF) is stable for a wide region of the phase diagram 
and almost completely veils the non-magnetic Mott transition for the unfrustrated case. 
However, AF is severely suppressed by the frustration, and the non-Magnetic Mott transition is realized for 
$t_1 \lesssim t_2$ (dimer side) even when the frustration is moderate ($0.3 \lesssim t_{3,4}/t_{1,2} \lesssim 0.5 $). 
In the region $t_2 \lesssim t_1$ (plaquette side), AF still veils the non-Magnetic Mott transition for moderate frustrations. 

{\it 1/5-depleted square lattice Hubbard model and VCA.}---

The Hamiltonian of the Hubbard model on the 1/5-depleted square lattice is given by 
\begin{eqnarray}
H =& -\sum_{i,j,\sigma} t_{ij}c_{i\sigma }^\dag c_{j\sigma}
+ U \sum_{i} n_{i\uparrow} n_{i\downarrow} - \mu \sum_{i,\sigma} n_{i\sigma},
\label{eqn:hm}
\end{eqnarray}
where $c^{(\dagger)}_{i\sigma}$ annihilates (creates) an electron with spin
$\sigma$ on the site $i$, $n_{i\sigma}=c^{\dagger}_{i\sigma}c_{i\sigma}$
, $U$ is the on-site Coulomb repulsion, and $\mu$ is the chemical potential. 
The hopping integrals $t_{i,j}=t_1$ ($t_2$) on plaquette- (dimer-) bonds, and 
$t_{i,j}=t_3$ ($t_4$) for the next-nearest-neighbor sites within (between) the plaquettes (See Fig. \ref{fig:model} (a)). 
We set the energy unit as $t_2 = 1$ ($t_1 = 1$) for $t_1/t_2 \leq 1.0$ ($t_2/t_1 \leq 1.0$). 
As for the effect of the frustration $t_{3,4}$, we introduce the frustration parameter $f = t_3/t_1 = t_4/t_2 $ 
and consider the three cases $f = 0$, $0.3$, and $0.5$. 
When unfrustrated ($f=0$), the system has a particle-hole symmetry at half-filling, and for noninteracting case ($U=0$) 
the ground state is a band insulator (metal) for $t_1/t_2 < 0.5$ ($0.5 < t_1/t_2$). 

We use VCA\cite{senechal00} in our analysis. 
In this approach, we write the thermodynamic potential of the system $\Omega _{\mathbf{t}}[\Sigma ]$ 
in the form of a functional of the self-energy $\Sigma $, which is stationary 
$\delta \Omega_{\mathbf{t}}[\Sigma ]/\delta \Sigma =0$ at the physical self-energy, and evaluate 
it for the exact self-energy of a simpler Hamiltonian $H'$ which shares the same interaction part with $H$ 
($\mathbf{t}$ stand for the explicit dependence of $\Omega _{\mathbf{t}}$ on all the one-body operators in the Hamiltonian). 
As for $H'$ we use the same model defined on the disconnected identical clusters 
(referred to as the reference clusters, hereafter), which tile the original infinite clusters, 
and add it various Weiss fields for analyzing symmetry breaking. 
Then the functional $\Omega _{\mathbf{t}}[\Sigma ]$ is reduced to the function of the one body operators $\mathbf{t}'$ of $H'$,
expressed as 
\begin{equation}
\Omega _{\mathbf{t}}(\mathbf{t}')=\Omega'\kern-0.4em - \kern-0.4em\int_C{\frac{%
d\omega }{2\pi }} {\rm e}^{ \delta \omega} \sum_{\mathbf{K}}\ln \det \left(
1+(G_0^{-1}\kern-0.2em -G_0'{}^{-1})G'\right),
\label{omega}
\end{equation}%
where $\Omega'$ and $G'$ are the thermodynamic potential and exact Green's function of $H'$, 
$G_0$ and $G_0'$ are the noninteracting Green's function of $H$ and $H'$, respectively, 
the frequency integral is carried along the imaginary axis $\delta \rightarrow + 0$ 
and the sum is over the superlattice wave vectors. 
The stationary solution of $\Omega_{\mathbf{t}}(\mathbf{t}')$ and the 
exact self-energy of $H'$ at the stationary point are the approximate grand-potential 
and self-energy of $H$ in VCA. 
In VCA, the short-range correlations within the reference cluster are exactly taken into account and 
the restriction of the space of the self-energies $\Sigma$ into that of $H'$ is the only approximation. 

In our analysis, the 8- and 12-site clusters in Fig.~\ref{fig:model}(b)-(c), which are referred to as (b) 8D and 12D, and 
(c) 8P and 12P hereafter, are used, and the Weiss field 
$H_{\rm AF}= h_{\rm AF}\sum_{i} {\rm sign}(i)(n_{i\uparrow }-n_{i\downarrow })$ 
with ${\rm sign}(i) = -1 (1)$ for the up (down) spin sites is included. 
In the stationary point search of $\Omega(\mu', h_{\rm AF})$, which we denote as the grand-potential per site, 
$h_{\rm AF}$ and the cluster chemical potential $\mu'$ are treated as the variational parameters, where 
the latter should be included for the thermodynamic consistency\cite{aichhorn}. 
During the search, the chemical potential of the system $\mu$ is also adjusted so that the electron 
density $n$ is equal to 1 within 0.1\%. 

On 8D and 12D (8P and 12P) every site can form a dimer (plaquette) with other site(s) on the same reference cluster. 
In VCA, the AF ordering is studied by including the Weiss field in the reference Hamiltonian and since the 
formation of the dimers or plaquettes is one of the leading competitors for the AF ordering, for example, 
the sites in 8P and 12P which can not form a dimer within the same reference cluster 
tend to magnetically order easily in the dimer side ($t_1 \lesssim t_2$) due to the AF Weiss field, 
which does not correspond to physical situations. 
Similar phenomena take place in the plaquette side ($t_2 \lesssim t_1$) on 8D and 12D. 
So we combine the results of the dimer- and plaquette-type clusters to circumvent the unphysical situations, 
and identify the AF phase as the region where AF is stable both on 12D and 12P. 
\begin{figure}[t]
\begin{center}
\includegraphics[width=8cm,bb= 145 96 463 542]{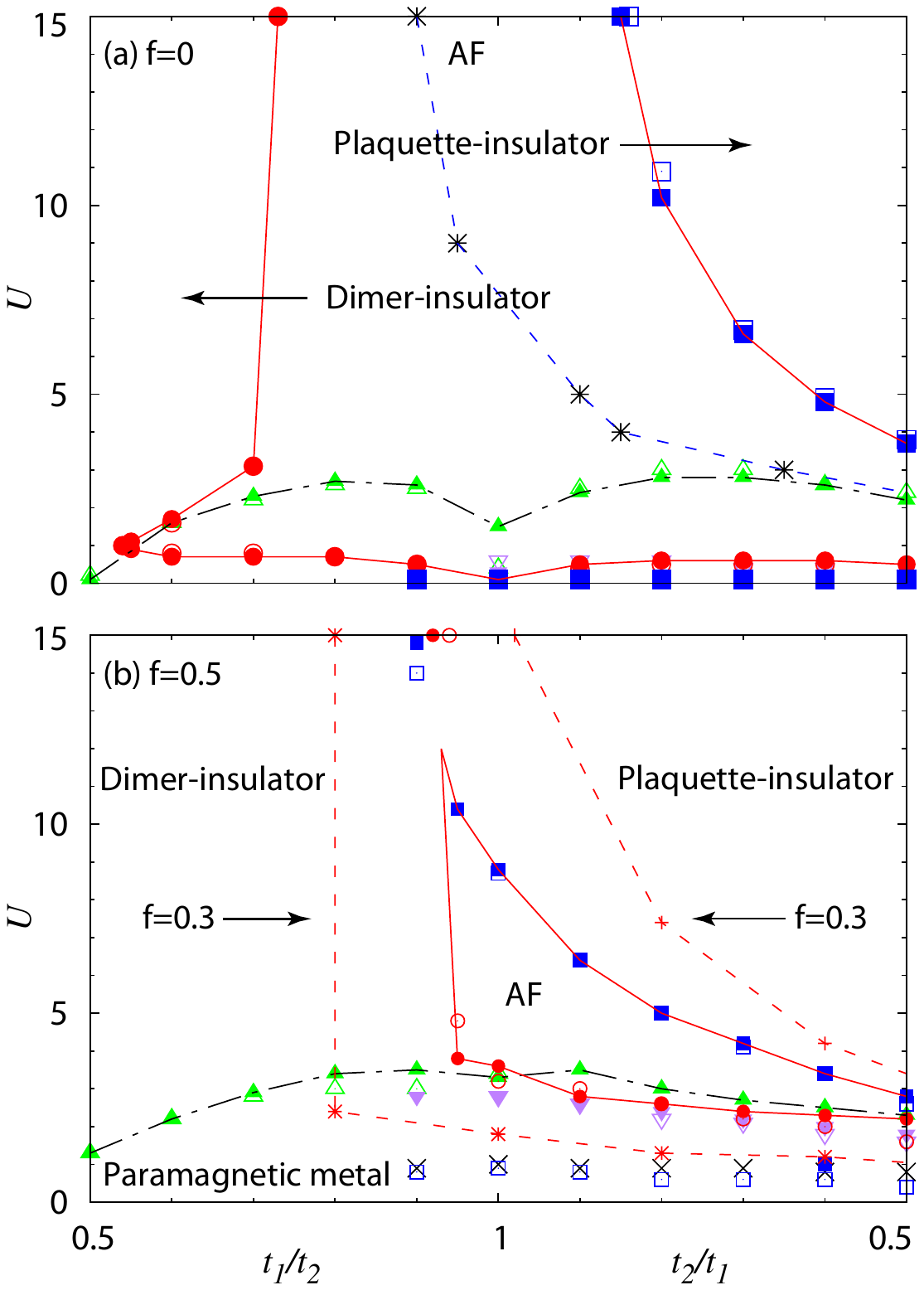}
\caption{
(Color online) Phase diagram of the Hubbard model on the 1/5-depleted square lattice with (a) $f = 0$ 
and (b) $f = 0.5$. The filled circles (squares) correspond to the magnetic transition points and 
the up (down) triangles correspond to the Mott transition points computed imposing $h_{\rm AF} = 0$ on 12D (12P). 
The unfilled marks correspond to the results of 8D and 8P clusters. Lines are guides to the eye. 
\label{fig:phase}}
\end{center}
\end{figure}

{\it Magnetic properties.}---
Fig.~\ref{fig:phase} shows the phase diagram with (a) $f = 0$ and (b) $f = 0.5$ computed by VCA. 
The filled circles (squares) correspond to the magnetic transition points $U_c$ and 
the up (down) triangles correspond to the Mott transition points $U_{\rm MI}$ computed with $h_{\rm AF} = 0$ 
on 12D (12P). The unfilled marks correspond to the results on 8D and 8P. 
In Fig.~\ref{fig:phase} (a) the asterisks denote the points where the magnetic order parameter $M$ takes its maximum 
as the function of $t_1$ ($t_2$) for fixed $U$. 
In Fig.~\ref{fig:phase} (b) the asterisks (pluses) correspond to $U_c$ at $f= 0.3$ computed on 12D (12P). 
The crosses denote the points below which the paramagnetic ground state on 12P 
becomes spin-triplet, and the magnetic properties below these points are not analyzed. 
On 12P the AF phase persists down to these points for $0.9 \le t_1/t_2$ except at $t_2/t_1 = 0.6$. 
Both the magnetic and non-magnetic Mott transitions are of the second order 
since there are no energetically disfavored AF, insulator, or metallic solutions 
around the transition points. 
Comparing Fig.~\ref{fig:phase} (a) and (b), AF is stable in a wide region and almost completely veils 
the non-magnetic Mott transition for $f=0$. 
However, AF is highly suppressed by the frustration, especially in the dimer side, and 
Mott transition is realized without being veiled by AF in this range. 
In the plaquette side, non-magnetic Mott transition is still veiled by AF even for the frustrated case on all the four clusters. 
The cluster size dependence of the results among the same type of clusters (between 8D and 12D, or 8P and 12P) is rather small. 
\begin{figure}[t]
\begin{center}
\includegraphics[width=8cm,bb=142 309 465 536]{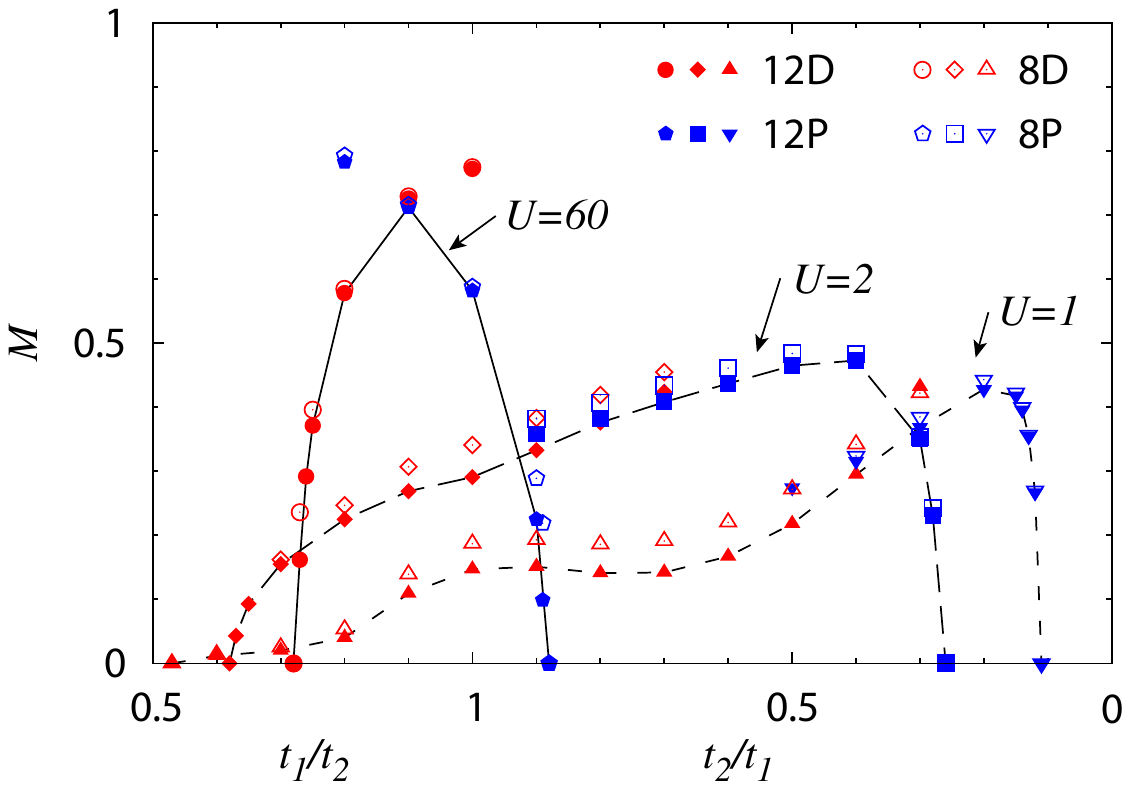}
\caption{
(Color online)
Magnetic order parameter $M$ with $f= 0$ at $U = 60$, $2$, and $1$ computed by VCA. 
Lines are guides to the eye, where the lower values of $M$ among 12D and 12P are connected. 
\label{fig:orderf0}}
\end{center}
\end{figure}
In Fig.~\ref{fig:phase} we connected the data of $U_{\rm MI}$ on 12D even in the plaquette side $1.0 \gtrsim t_2/t_1$. 
We discuss on this point later with the analysis of $\langle S_i \cdot S_j \rangle$ on 12D and 12P, 
however, our overall conclusions are not affected by this preference since AF veils the Mott transition 
on all the four clusters in the plaquette side. 
\begin{figure}[t]
\begin{center}
\includegraphics[width=9cm,bb=96 195 535 490]{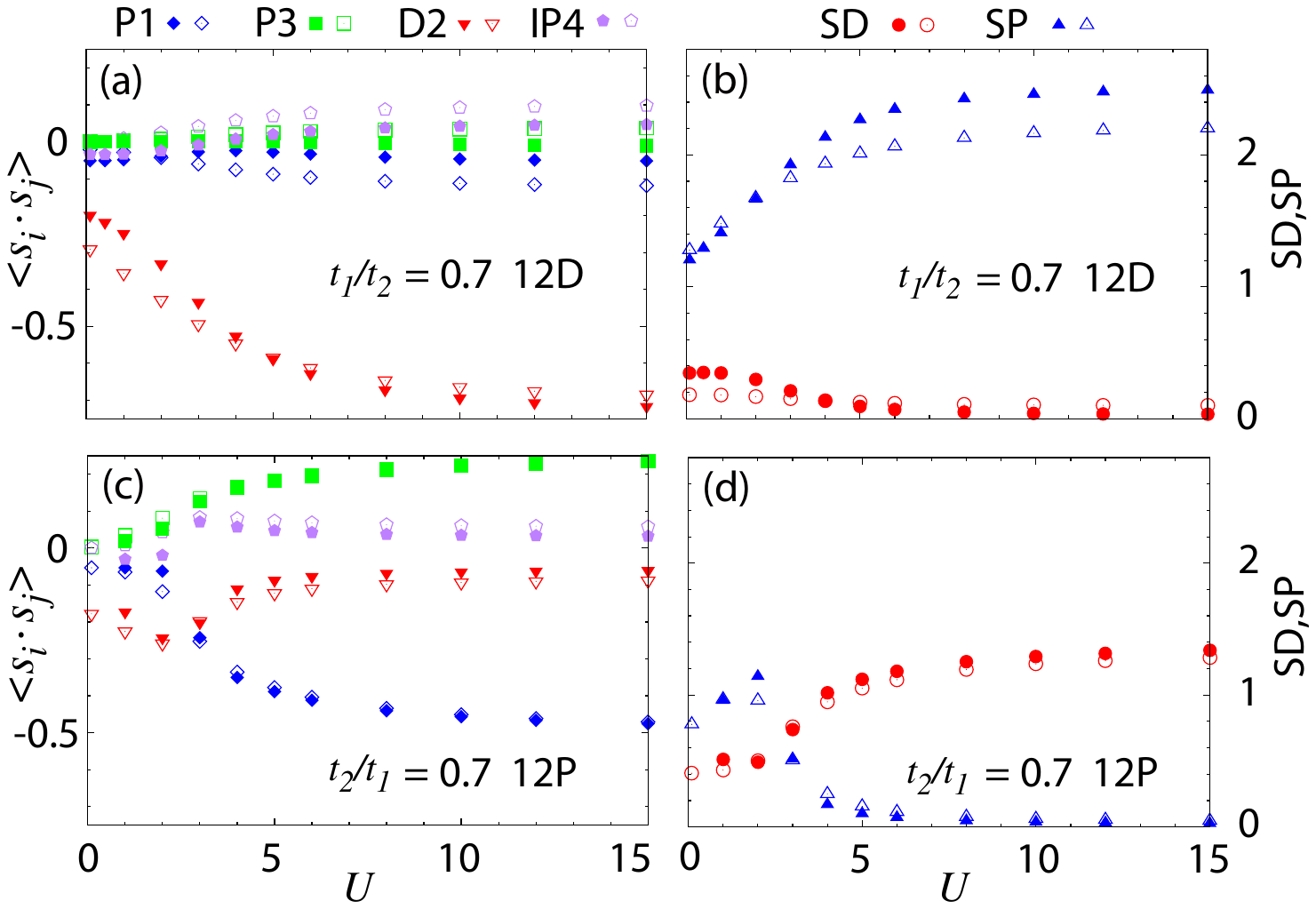}
\caption{
(Color online) 
The spin spin correlation $<S_i \cdot S_j>$ on different links 
of the reference clusters computed assuming 
$h_{\rm AF} = 0$ (a) at $t_1 =0.7$ on 12D cluster (b) at $t_2=0.7$ on 12P cluster.
\label{fig:ss}}
\end{center}
\end{figure}
Fig.~\ref{fig:orderf0} shows the magnetic order parameter $M$ with $f= 0$ at $U = 60$, $2$, and $1$ 
computed by VCA. Lines are guides to the eye, where the lower values of $M$ among 12D and 12P are connected. 
After combining the 12D and 12P data, the position of the maximum of $M$ shifts from the dimer side $t_1/t_2 \simeq 0.9$ 
to the plaquette side $1 > t_2/t_1 $ as $U$ decreases. 

Here we briefly compare our results for the unfrustrated case $f=0$ with other analyses. 
Comparing with the Heisenberg analyses, our critical hoppings $t_{1,2c}$ separating the AF and non-magnetic phase are 
$t_{1c} = 0.73$ $(0.73)$ and $t_{2c} = 0.89$ $(88)$ for $U = 60 (30)$, while the Heisenberg results\cite{troyer} are 
$0.74 < t_{1c} < 0.81$ and $0.96 < t_{2c} < 0.98$. 
The magnetic transition is of the second order and order parameter $M$ takes its maximum at $t_1/t_2 \simeq 0.9$ 
both by the Heisenberg model\cite{troyer} and VCA with $U=60$. 
Therefore the results of the Heisenberg model model and VCA with $U=60$ agree very well except the value of $t_{2c}$. 

Comparing with the CDMFT\cite{yanagi}, our $U_{MI}$ is smaller than that of CDMFT. 
This tendency is observed also e.g., on another non-Brave lattice: the kagom\'e lattice\cite{kawakami10,atsushi2011}. 
The Mott transition is of the second order both in VCA and CDMFT for $f=0$ where the particle-hole symmetry holds at half-filling. 
In general, on other lattices, VCA predicts the second order Mott transition\cite{atsushi2011,atsushi2013,atsushi2014} 
while the cluster mean field theories with bath degrees of freedom predict the first order phase transition\cite{Potthoff2}. 

Comparing with the DQMC\cite{khatami}, our $U_c$ is very similar to that of DQMC, 
though only the three values of $U$ are analyzed in DQMC. 
In DQMC the high symmetry point shifts from the dimer side to the plaquette side as $U$ decreases.   

Our analysis is the first one studying the effect of the frustration in the Hubbard model, 
and only the Heisenberg results are available for frustrated cases $f \ne 0$. 
In Fig.~\ref{fig:phase} (b) AF is realized with $f=0.3$ and non-magnetic insulator is realized with $f=0.5$ at $t_1/t_2 =1$ 
for $U=15$, and this is quantitatively consistent with the results of Ref. \onlinecite{albrecht}. 

To further study the nature of the phases, in Fig.~\ref{fig:ss} (a)$\sim$(d) we show 
the correlation $\langle S_i \cdot S_j \rangle$ on different links: 
Along the side of the plaquette $\langle S_1 \cdot S_2 \rangle$ (P1), 
along the diagonal of the plaquette $\langle S_1 \cdot S_3 \rangle$ (P3), 
along the dimer $\langle S_1 \cdot S_5 \rangle$ (D2), 
and along the $t_4$-direction $\langle S_1 \cdot S_6 \rangle$ (IP4), 
together with the total spin squared of the dimer $\langle (S_1 + S_5 )^2 \rangle$ ${\rm SD}$ 
and that of the plaquette ($\langle (\Sigma_{i=1}^{4} S_i )^2 \rangle$  (${\rm SP}$))
(See Fig. \ref{fig:model} (b) and (c) for the position of the site $1\sim 7$)
computed by the exact diagonalization of 12D for $t_1/t_2 =0.7$ and 12P for $t_2/t_1 =0.7$ 
at the stationary non-magnetic solutions of VCA. 
The filled (unfilled) marks correspond to $f=0.5$ ($f=0$).   
The values of P1, P3, D2, IP4, ${\rm SD}$, and ${\rm SP}$ in the pure dimer (plaquette) state are: 
$0$, $0$, $-0.75$, $0$, $0$, $3$ ($-0.5$, $0.25$, $0$, $0$, $1.5$, $0$). 
For $ 4 \lesssim U $, dimer (plaquette) state is realized at $t_1/t_2 =0.7$ ($t_2/t_1 =0.7$), and the frustration enhances the 
formations of the dimers (plaquettes) since SD (SP) is lower for frustrated case. 
For $ U \lesssim 2 \sim3 $, $\langle S_i \cdot S_j \rangle$ rapidly decreases and the frustration enhances this decrease, 
probably because of the enhancement of the mobility of the electrons or kinetic energies confirmed 
by the increase of $U_{MI}$ due to the frustration. 
For $ U \lesssim 2 \sim3 $, SP becomes larger than SD even in the plaquette side on 12P 
and we confirmed that this behavior is observed also on 12D in the plaquette side. 
This reverse will be because, since the maximum value of the total spin squared realized on the 4-site plaquette is 
$2\times(2+1)$ and is larger than that of the 2-site dimer $1\times(1+1)$,
as $U$ decreases and the plaquette state begins to melt, SP becomes larger than SD even in the plaquette side. 
This tendency will be too exaggerated on 8P and 12P, which are filled with plaquettes, 
and this feature leads to the lower $U_{c}$ and the spin triplet non-magnetic ground states for small $U$ on 12P. 
So the results of 12D will be better approximation of the infinite system in the plaquette side 
$1.0 \gtrsim t_2/t_1$ for $U$ lower than $U_{MI}$ of 12D. 
 
{\it Summary and discussion.}---
In summary, we have studied the magnetic phase diagram and Mott transition in 
the half-filled Hubbard model on the 1/5-depleted square lattice by VCA, taking into account the effect of the 
frustrations. 
We found that the $(\pi,\pi)$ N\'eel ordering AF is stable in a wide region of the phase diagram
and almost completely veils the non-magnetic Mott transition for the unfrustrated case. 
However, the AF is severely suppressed by the frustration and even with moderate frustrations $f \simeq 0.3 \sim 0.5$ 
non-magnetic Mott transition takes place in the range $t_1/t_2 \lesssim 0.8 \sim 1.0$. 
The nature of the non-magnetic insulator is very close to the dimer-insulator (plauette-insulator) above the magnetic or Mott transition points.  

\begin{acknowledgments}
The author would like to thank T.~Inakura, H.~Kurasawa and H.~Nakada 
for useful discussions on the numerical analysis. Parts of numerical calculations were done using the computer facilities of 
the IMIT at Chiba University and Yukawa Institute. 

\end{acknowledgments}


\end{document}